\documentclass[aps, prb, twocolumn]{revtex4}

\usepackage{amstext}
\usepackage{amsmath}
\usepackage{amssymb}
\usepackage{amsfonts}
\usepackage{graphicx}
\usepackage[hang,nooneline, scriptsize,tight]{subfigure}

\begin{document}

\title{Decoherence and Quantum-Classical Master Equation Dynamics}
\author{Robbie Grunwald and Raymond Kapral}
\affiliation{Chemical Physics Theory Group, Department of
Chemistry, University of Toronto, Toronto, ON M5S 3H6 Canada}

\date{\today}

\begin{abstract}
The conditions under which quantum-classical Liouville dynamics may be reduced to a master equation are investigated.  Systems that can be partitioned into a quantum-classical subsystem interacting with a classical bath are considered.  Starting with an exact non-Markovian equation for the diagonal elements of the density matrix, an evolution equation for the subsystem density matrix is derived.  One contribution to this equation contains the bath average of a memory kernel that accounts for all coherences in the system.  It is shown to be a rapidly decaying function, motivating a Markovian approximation on this term in the evolution equation. The resulting subsystem density matrix equation is still non-Markovian due to the fact that bath degrees of freedom have been projected out of the dynamics.  Provided the computation of non-equilibrium average values or correlation functions is considered, the non-Markovian character of this equation can be removed by lifting the equation into the full phase space of the system.  This leads to a trajectory description of the dynamics where each fictitious trajectory accounts for decoherence due to the bath degrees of freedom. The results are illustrated by computations of the rate constant of a model nonadiabatic chemical reaction.
\end{abstract}

\maketitle

\section{Introduction} \label{sec:intro}

When investigating quantum relaxation processes in the condensed phase, one often partitions the full quantum system into a subsystem whose dynamics is of interest and an environment or bath with which the subsystem interacts.  There is a large literature dealing with such open quantum systems.~\cite{weiss99, breuer02} A number of different equations of motion for the density matrix of the subsystem have been derived, including the Lindblad~\cite{lindblad76} and Redfield~\cite{redfield57} equations and a variety of generalized quantum master equations.~\cite{caldeira-leggett, unruh:1989, hu92, karrlein:1997, romero-rochin89, shi:2003, esposito:2003}  Effects due to the environment typically enter these equations through coupling terms involving parameters that characterize the bath relaxation processes.  Such equations have been used to investigate aspects of decoherence in the quantum subsystem arising from interactions with bath degrees of freedom.    

Sometimes it is convenient to suppose that the dynamics of certain degrees of freedom is described by quantum mechanics while other degrees of freedom may be treated by classical mechanics to a good approximation.  This is the case if one considers systems involving light particles interacting with a bath of heavy particles.  Proton and electron transfer processes in the condensed phase and in biomolecules fall into this category, as do many vibrational relaxation processes.  One is then led to study the dynamics of quantum-classical systems where the entire system is partitioned into quantum and classical subsystems.~\cite{kapral:2006}  Equations of motion for the quantum subsystem density matrix, where the classical bath is modeled as a dissipative environment have been derived.~\cite{kapral&toutounji01, toutounji05, toutounji06} Such descriptions are useful for many applications; however, there are situations where the quantum subsystem evolution depends explicitly on the details of the bath dynamics.  This is important since specific features of bath motions can influence quantum rate processes.~\cite{com1}  To describe such specific bath dynamical effects one must use the full quantum-classical equation of motion.  

Another partition of the system is required when the quantum subsystem is directly coupled to a small subset of the environmental degrees of freedom.  For example, in proton transfer within a bio-molecule, the quantum subsystem may be taken to be the proton, which interacts directly with a specific set of functional groups of a larger molecule immersed in a solvent.  In this case, we may define a quantum-classical subsystem comprising both quantum degrees of freedom (the proton) and a subset of the classical variables that directly couple to these quantum degrees of freedom (the specified functional groups).  The remaining classical variables constitute the bath.  In such a partition, the bath may be treated either explicitly or as a dissipative environment.  Equations of motion for the density matrix of a quantum-classical subsystem interacting with a dissipative bath have been derived.~\cite{kapral01}  In this article we consider quantum-classical systems of this type but instead retain the details of the bath dynamics and study the conditions under which the dynamics can be reduced to a master equation.

The reduction of quantum-classical dynamics to a subsystem master equation hinges on the decoherence in the subsystem induced by interactions with the bath degrees of freedom.~\cite{com3}  Consequently, we focus on how decoherence is described in quantum-classical systems and the conditions under which it is strong enough to eliminate the off-diagonal elements of the density matrix on short time scales. The result of the analysis is a non-Markovian generalized master equation for the density matrix of the quantum-classical subsystem.  When this equation is used to compute non-equilibrium averages or correlation function expressions involving subsystem properties, we show that the subsystem dynamics may be lifted to the full phase space, including the bath degrees of freedom, to recover a Markovian master equation for the quantum and all classical degrees of freedom.  Solutions of this equation may be obtained from an ensemble of surface-hopping trajectories, each member of which incorporates the effects of quantum decoherence.

The outline of the paper is as follows:  The starting point of our analysis is the quantum-classical Liouville equation for the entire system.~\cite{gerasimenko:1981, geras82, alek81, boucher88, zhang88, martens97, horenko02, shi04, kapral99, kisil:2005, prezhdo:2006}  In Sec.~\ref{sec:genmaster} we show how this equation can be cast into the form of a generalized master equation for the diagonal elements of the density matrix.  This equation involves a memory kernel operator that contains all information on quantum coherence in the system.  We discuss the explicit form of the memory kernel operator that governs the evolution in off-diagonal space and present its functional form.  The analysis in Sec.~\ref{sec:RME} and Appendix~\ref{appendix:B} shows that an average over an ensemble of trajectories evolving on coherently coupled surfaces with different bath initial conditions can be used to reduce the \textit{full-system} generalized master equation to a non-Markovian \textit{subsystem} generalized master equation. We show that this equation can then be lifted back to the full phase space to obtain a Markovian master equation.  We apply this formalism in Sec.~\ref{sec:res} and calculate the nonadiabatic rate constant for a model system.  Numerical results using full quantum-classical Liouville dynamics and master equation dynamics are compared. In the concluding section we comment on the relationship of our master equation dynamics to other surface-hopping methods.

\section{Generalized Master Equation} \label{sec:genmaster}

Starting from the quantum-classical Liouville equation, it is not difficult to derive a generalized master equation for the diagonal elements of the density matrix. As described in the Introduction, the position and momentum operators, $(\hat{q}, \hat{p})$, of the quantum degrees of freedom are assumed to be coupled directly to a set of classical phase space variables, $X_0\equiv (R_0, P_0)$. Together these make up the quantum-classical subsystem.  The classical $X_0$ variables are, in turn, directly coupled to the remainder of the classical phase space variables, $X_b\equiv (R_b, P_b)$, that constitute the bath. (Our formulation must be modified If the quantum degrees of freedom couple directly to all other variables in the system.)  The total Hamiltonian of the system is,
\begin{eqnarray}
   \hat{H}(X) & = & \frac{P_b^2}{2M} + \frac{P_0^2}{2M} + \frac{\hat{p}^2}{2m} + \hat{V}(\hat{q},R_0, R_b) \nonumber \\
   & \equiv & \frac{P_b^2}{2M} + \frac{P_0^2}{2M} + \hat{h}(R) \;.
\end{eqnarray}
The potential energy operator, $\hat{V}(\hat{q}, R_0, R_b)$, includes the contributions from the quantum-classical subsystem, the bath, and the interaction between the two.  It is often convenient to represent the dynamics in the adiabatic basis given by, $\hat{h}(R_0, R_b) |\alpha;R_0 \rangle=E_\alpha (R_0, R_b) |\alpha;R_0 \rangle$, where $\hat{h}(R_0, R_b)$ is the quantum Hamiltonian for a fixed configuration of the classical particles.  The adiabatic eigenfunctions depend only on the coordinates $R_0$ since the  dependence on the bath coordinates enters through the potential as an additive constant.  In this basis, the equation of motion for the full density matrix is~\cite{kapral99},
\begin{equation}
   \frac{\partial }{\partial t}\rho_W^{\alpha \alpha'}(X, t) = - \sum_{\beta \beta'} i \mathcal{L}_{\alpha \alpha', \beta \beta'} \rho_W^{\beta \beta'}(X,   t), \label{Q-C Liouville}
\end{equation}
where $\rho_W^{\alpha \alpha'} (X, t)$ is a density matrix element which depends on the full set of phase space coordinates $X =(R,P) \equiv (X_0, X_b)$. The quantum-classical Liouville super-operator is given by,~\cite{kapral99}
\begin{equation}
    -i\mathcal{L}_{\alpha \alpha', \beta \beta'} = -i (\omega_{\alpha \alpha'} + L_{\alpha \alpha'}) \delta_{\alpha \beta} \delta_{\alpha' \beta'} + \mathcal{J}_{\alpha \alpha', \beta \beta'}\;,\label{Q-C superoperator}
\end{equation}
where $\omega_{\alpha \alpha'}=\Delta E_{\alpha \alpha'}/\hbar$, with $\Delta E_{\alpha \alpha'} = E_{\alpha} - E_{\alpha'}$, and the classical Liouville operator, $iL_{\alpha \alpha'}$ is given by
\begin{equation}
   iL_{\alpha \alpha'} = \frac{P}{M}\cdot{\partial \over \partial R} +\frac{1}{2} \left( F^{\alpha} + F^{\alpha'}
   \right) \cdot{\partial \over \partial P}\;. \label{classical Liouville}
\end{equation}
The Hellmann-Feynman force~\cite{feynman39} for state $\alpha$ is $F^{\alpha} = \langle \alpha ; R_0 | \partial \hat{V}(\hat{q}, R_0, R_b) / \partial R | \alpha ; R_0 \rangle$, and the operator $\mathcal{J}_{\alpha \alpha', \beta \beta'}$, defined in the next section, accounts for quantum transitions and corresponding momentum changes in the environment.

We shall often simplify the notation in what follows.  Above, $\rho_W (X, t)$ refers to the partially Wigner transformed density matrix whose matrix elements are $\rho_W^{\alpha \alpha'}(X,t)$.  Since we use partially Wigner transformed variables throughout this article, we shall drop the subscript W.  

The quantum-classical Liouville evolution operator can be partitioned into diagonal, off-diagonal and coupling components by defining the super-operators: $\mathcal{L}^d$, $\mathcal{L}^{d,o}$, $\mathcal{L}^{o,d}$, and $\mathcal{L}^o$, where the d, and o superscripts denote diagonal and off-diagonal, respectively.  We define the diagonal part of the density matrix $\rho_d (X, t)$, with matrix elements, $\rho^{\alpha \alpha}(X, t) \delta_{\alpha \alpha'} \equiv \rho_d^{\alpha} (X, t)$.  Similarly, the off-diagonal part of the density matrix, $\rho_o (X, t)$, has matrix elements $\rho^{\alpha \alpha'} (X, t) (1 - \delta_{\alpha \alpha'}) \equiv \rho_o^{\alpha \alpha'} (X, t)$. Using these definitions, the quantum-classical Liouville equation may be expressed formally as the following set of coupled differential equations,
\begin{eqnarray}
    \frac{\partial}{\partial t} \rho_d (X, t) & = & -i \mathcal{L}^d \rho_d (X, t) - i \mathcal{L}^{d,o} \rho_o (X, t) \label{diagonal evolution} \\
    \frac{\partial}{\partial t} \rho_o (X, t) & = & -i \mathcal{L}^{o} \rho_o (X, t) - i \mathcal{L}^{o, d} \rho_d (X, t). \label{offdiagonal evolution}
\end{eqnarray}
By substituting the formal solution of Eq.~(\ref{offdiagonal evolution}) into Eq.~(\ref{diagonal evolution}), we obtain the
evolution equation for $\rho_d(X,t)$,
\begin{eqnarray}
   \frac{\partial}{\partial t} \rho_d (X, t) & = & - i \mathcal{L}^{d, o} e^{-i \mathcal{L}^o t} \rho_o (X, 0) - i \mathcal{L}^d \rho_d (X, t) \label{projected evolution} \\
   & & + \int_0^t dt' i \mathcal{L}^{d, o} e^{-i \mathcal{L}^o (t - t')} i \mathcal{L}^{o, d} \rho_d (X, t') \nonumber
\end{eqnarray}
For the remainder of this analysis we will assume that $\rho_o(X,0) = 0$, and thus the first term vanishes.  This amounts to initially preparing the system in a pure state or incoherent mixture of states.  Although equation~(\ref{projected evolution}) is general and may be used to study systems that are initially prepared in coherent states, we shall not consider such situations here.  Using Eq.~(\ref{Q-C superoperator}), the explicit definitions of the matrix elements of the Liouville super-operators are,
 \begin{eqnarray}
   i \mathcal{L}^d_{\alpha \alpha', \beta \beta'} & \equiv & i L_{\alpha} \delta_{\alpha \beta} \delta_{\alpha \alpha'} \delta_{\beta \beta'}\nonumber \\
   i \mathcal{L}^{d,o}_{\alpha \alpha', \beta \beta'} & \equiv & - \mathcal{J}^{d,o}_{\alpha, \beta \beta'} \delta_{\alpha \alpha'}(1 - \delta_{\beta \beta'}) \label{liouville-matrix-elements} \\
   i \mathcal{L}^{o,d}_{\alpha \alpha', \beta \beta'} & \equiv & - \mathcal{J}^{o,d}_{\alpha \alpha', \beta} (1 - \delta_{\alpha \alpha'}) \delta_{\beta \beta'}\nonumber \\
   i \mathcal{L}^o_{\alpha \alpha', \beta \beta'} & \equiv & i \mathcal{L}_{\alpha \alpha', \beta \beta'} (1 - \delta_{\alpha \alpha'}) (1 - \delta_{\beta \beta'}), \nonumber
\end{eqnarray}
where $i L_{\alpha}=i L_{\alpha \alpha}$ and $\mathcal{J}^{d,o}_{\alpha, \beta \beta'} = \mathcal{J}_{\alpha
\alpha, \beta \beta'}$, for $\beta \neq \beta'$, with a similar definition for $\mathcal{J}^{o,d}$.  Using these definitions and the initial condition discussed above, we obtain the generalized master equation for the evolution of the diagonal elements of the density matrix.
\begin{eqnarray}
   \frac{\partial}{\partial t} \rho_d^{\alpha} (X, t) & = & -i L_{\alpha} \rho_d^{\alpha} (X, t) \nonumber \\
   & & + \int_0^t dt' \sum_{\beta} \mathcal{M}_{\alpha \beta} (t') \rho_d^{\beta} (X, t - t') \;,\label{gme}
\end{eqnarray}
where the memory kernel operator $\mathcal{M}_{\alpha \beta} (t)$ is given by,
\begin{equation}
   \mathcal{M}_{\alpha \beta} (t) = \sum_{\nu \nu' , \mu \mu'} \mathcal{J}^{d,o}_{\alpha, \mu \mu'} \left( e^{-i \mathcal{L}^o (X) (t)}\right)_{\mu \mu', \nu \nu'} \mathcal{J}^{o,d}_{\nu \nu', \beta} \;, \label{mem kern}
\end{equation}
and acts on all of the classical degrees of freedom that appear in functions to its right. Next, we analyze the form of the memory kernel operator~(\ref{mem kern}) in order to cast it into a form that is suitable for the derivation of a master equation.

\subsection*{Memory kernel} \label{sec:memkern}

The explicit form of the $\mathcal{J}$ operator was derived previously~\cite{kapral99} and is given by
\begin{eqnarray}
   \mathcal{J}_{\alpha \alpha', \beta \beta'} & = & \mathcal{C}_{\alpha \beta} \delta_{\alpha' \beta'} + \mathcal{C}^*_{\alpha' \beta'} \delta_{\alpha \beta}, \label{Joperator}
\end{eqnarray}
where
\begin{eqnarray}
   \mathcal{C}_{\alpha \beta} & = & -D_{\alpha \beta}(X_0) \left(1 + \frac{1}{2} S_{\alpha \beta}\cdot \frac{\partial}{\partial P_0} \right) \;, \label{Cop}
\end{eqnarray}
the nonadiabatic coupling matrix element is given by $d_{\alpha \beta} = \langle \alpha ; R_0 |\nabla_{R_0} | \beta ; R_0 \rangle$, $S_{\alpha \beta} =\Delta E_{\alpha \beta} d_{\alpha \beta}/D_{\alpha \beta}(X_0)$, and $D_{\alpha \beta}(X_0) = (P_0/M_0) \cdot d_{\alpha \beta}$.  We have shown in earlier work that the action of the $\mathcal{C}$ operator on phase space functions may be computed using the momentum-jump approximation~\cite{kapral&ciccotti01, hanna05}
\begin{eqnarray}
   \mathcal{C}_{\alpha \beta}(X_0) & \approx & -D_{\alpha \beta}(X_0) j_{\alpha \beta}(X_0)\;,
\end{eqnarray}
where the momentum shift operator, $j_{\alpha \beta} (X_0)$, is a translation operator in momentum space,
\begin{eqnarray}
   j_{\alpha \beta} f(P_0) & \equiv & e^{\Delta E_{\alpha \beta} M_0 \partial / \partial (P_0\cdot \hat{d}_{\alpha \beta})^2} f(P_0) \nonumber \\
   & = & f(P_0+\Delta {P_0}_{\alpha \beta}), \label{full-mom-jump} \\
   \Delta {P_0}_{\alpha \beta} & = & \hat{d}_{\alpha \beta} \Big({\rm sgn}(P_0 \cdot \hat{d}_{\alpha \beta})
   \label{eq:deltaP} \\
   & &  \times \sqrt{( P_0 \cdot \hat{d}_{\alpha \beta})^2 + \Delta E_{\alpha \beta}M_0} - (P_0 \cdot \hat{d}_{\alpha \beta})\Big) .  \nonumber
\end{eqnarray}
Since momentum shifts occur in conjunction with quantum transitions, they depend on the quantum states involved in the transition.  Consequently, we use the following notation: $\bar{X}_{0 \alpha \beta} = (R_0, \bar{P}_{0 \alpha \beta}) = (R_0, P_0 + \Delta{P_0}_{\alpha \beta})$, so that, $j_{\alpha \beta}(X_0) f(X_0) = f (\bar{X}_{0 \alpha \beta})$.  It is worth noting here that the momentum shift operators do not act on the full classical environment.  They only act on the classical variables $X_0$ that directly couple to the quantum degrees of freedom. We also observe that the argument of the square root in Eq.~(\ref{eq:deltaP}) must be positive.  This condition prevents quantum transitions when there is insufficient energy in the classical degrees of freedom to effect the transition.

The time evolution in Eq.~(\ref{mem kern}), is given by the propagator, $ e^{-i \mathcal{L}^o (X) t}$.  The quantum-classical Liouville operator in this expression acts on the entire phase space $X$ and accounts for the following processes: classical evolution of the bath coordinates $X_b$ and evolution of the classical subsystem coordinates $X_0$ on the mean potential surface $(E_\mu +E_\mu')/2$ with an associated phase factor. This evolution is interspersed with quantum transitions taking the subsystem to other coherently coupled states where evolution is again on mean surfaces with associated phase factors.  In the course of this evolution the system never returns to a diagonal state involving evolution on a single adiabatic surface.  For this reason we refer to such evolution as being in ``off-diagonal space".

If one considers the explicit action of the propagators appearing in the memory kernel, one can show how this operator acts on an arbitrary function of the phase space coordinates $X$. The details are given in Appendix~\ref{appendix:A}. Using the results obtained there, one can show that the generalized master equation can be written as,
\begin{eqnarray}
   &&   \frac{\partial }{\partial t}\rho_d^{\alpha} (X, t) = - i L_{\alpha} \rho_d^{\alpha} (X, t) \label{meqn-wc} \\
   && \qquad + \int_0^t dt' \Big( \sum_{\beta} M_{\alpha \beta}^{\alpha \beta}(X,t') \rho_d^\beta(\bar{X}^{\alpha \beta}_{0 \alpha \beta,t'}, X_{b,t'}, t-t') \nonumber \\
   & & \quad \qquad + \sum_{\nu}M_{\alpha \nu}^{\nu \alpha}(X,t')\rho_d^\alpha (\bar{X}^{\nu \alpha}_{0\alpha \nu , t'}, X_{b,t'}, t-t') \Big) \;. \nonumber
\end{eqnarray}
In this expression the superscripts on the coordinates indicate the action of a second momentum shift operator, ($j_{\nu \alpha}(\bar{X}_{0 \alpha \nu }) f(\bar{X}_{0 \alpha \nu}, X_b) = f(\bar{X}^{\nu \alpha}_{0\alpha \nu }, X_b)$).  This result provides us with a definition of the memory {\em function}, 
\begin{equation} \label{mk_function}
   M_{\alpha \beta}^{\alpha \beta}(X,t) = 2{\rm Re} \Big[ \mathcal{W}_{\alpha \beta}(t',0)\Big] D_{\alpha \beta}(X_0) D_{\alpha \beta}(\bar{X}_{0 \alpha \beta, t'}),
\end{equation}
where the subscripts and superscripts on the memory function label the indices on the first and second $D$ function respectively.  The phase factor $\mathcal{W}_{\alpha \beta}$ is defined as
\begin{eqnarray}
   \mathcal {W}_{\alpha \beta}(t_1,t_2)=  e^{-i\int_{t_1}^{t_2} d\tau \; \omega_{\alpha \beta}(R_{0\alpha \beta, \tau})}  \;. \label{eq:wphase1}
\end{eqnarray}
Now the actions of all classical propagators and momentum jumps have been accounted for explicitly in Eq.~(\ref{meqn-wc}) so that the memory kernel is a function of the phase space variables.

Given the assumption about the initial condition on the density matrix ($\rho_o(X, 0)=0$), for a two-level system the generalized master equation~(\ref{meqn-wc}) is fully equivalent to the quantum-classical Liouville equation from which it was derived.  This result is also applicable to multi-level systems in a weak coupling limit where quantum transitions among different coherently coupled states are neglected in the the off-diagonal propagator.  This amounts to neglecting terms higher than quadratic order in the nonadiabatic coupling strength, $d_{\alpha \beta}$, in the evolution operators.  The time evolution described by Eq.~(\ref{meqn-wc}) consists of classical evolution along single adiabatic surfaces and two memory terms.  The memory terms account for transitions to the mean surface, evolution along this surface, and transitions to a new adiabatic surface with rate $M_{\alpha \beta}^{\alpha \beta}$, or transitions back to the original surface with rate $M_{\alpha \nu}^{\nu \alpha}$.  Thus, the dynamics of the generalized master equation derived here is separated into diagonal and off-diagonal components providing a framework within which to investigate decoherence in the quantum-classical subsystem induced by the bath.

\section{Master Equation}\label{sec:RME}

We next consider the conditions under which the generalized master equation~(\ref{meqn-wc}) may be reduced to a simple master equation without memory.  This reduction hinges on the ability to consider the memory kernel as a rapidly decaying function so that a Markovian approximation can be made.  From its form in Eq.~(\ref{mk_function}) one can see that $M (X,t)$, which contains all information on quantum coherence, is an oscillatory function.  As a result, a Markovian approximation to the memory kernel cannot be made directly on the full phase space equation since there is no obvious mechanism for the decay of the memory function.  It is the decoherence by the environment that provides such a mechanism.

In this analysis we exploit the fact that decoherence has its origin in interactions with the bath degrees of freedom.  We have already observed that we are interested in dynamical properties of the quantum-classical subsystem.  For instance, nonequilibrium average values of interest have the form,
\begin{eqnarray}
   \overline{A(t)} & = & \sum_{\alpha \beta} \int dX_0 \int dX_b A^{\beta \alpha}(X_0) \rho^{\alpha \beta}(X, t) \nonumber \\
   & = & \sum_{\alpha \beta} \int dX_0 A^{\beta \alpha}(X_0) \rho^{\alpha \beta}_s (X_0, t) \;, \label{reduced-density}
\end{eqnarray}
where $A^{\beta \alpha}(X_0)$ are the matrix elements of a property of the subsystem and $\rho^{\alpha \beta}_s (X_0, t) \equiv \int dX_b \rho^{\alpha \beta}(X, t)$ is the subsystem density matrix.  If the operator $A^{\beta \alpha}(X_0)$ is diagonal, then only the diagonal elements of the subsystem density matrix are needed to compute its average value.  Alternatively, if decoherence quickly destroys the off-diagonal subsystem density matrix elements, then, after a short transient, only the diagonal elements will be needed to compute the expectation value. Later, we shall show that similar considerations can be used to evaluate correlation function expressions for transport properties of the subsystem.

To compute such average quantities, we see that we need the subsystem density matrix elements.  Starting with the generalized master equation in full phase space (Eq.~(\ref{meqn-wc})), we can introduce a bath projection operator,~\cite{com2}
\begin{eqnarray}
   \mathcal{P} \cdot & = & \rho_b^{c}(X_b; R_0) \int dX_b \cdot \;,\label{p-oper}
\end{eqnarray}
whose action on the density matrix yields the subsystem density matrix.  Here $\rho_b^{c}(X_b; R_0)$ is the bath density conditional on the subsystem configuration space variables $R_0$.  The bath density function is in general quantum mechanical but in some applications it may be replaced by its high temperature classical limit.  In the projection operator formalism it is not necessary to distinguish between these two cases.  The complement of $\mathcal{P}$ is $\mathcal{Q}$. In Appendix~\ref{appendix:B}, using standard projection operator methods,~\cite{zwanzig61} we argue that bath averaged correlations involving the fluctuations of the memory kernel from its bath average may be neglected.  In this case, the subsystem generalized master equation is,
\begin{eqnarray}
   & & \frac{\partial}{\partial t} \rho_s^\alpha (X_0,t) = \label{penult-PRME-folded} \\
   & & \quad -\langle i L_\alpha  e^{-i\mathcal{Q}L_\alpha t} \mathcal{Q}\rho_d^\alpha (X, 0) \rangle_b - \langle i L_\alpha \rangle_b \rho_s^\alpha(X_0,t) \nonumber \\
   & & \quad - \int_0^t dt' \langle i  L_\alpha e^{- i \mathcal{Q} L_\alpha t'} i \mathcal{Q} L_\alpha \rangle_b \rho_s^\alpha (X_0, t - t') \nonumber \\
   && \quad + \int_0^t dt' \Big( \sum_{\beta} \langle M_{\alpha \beta}^{\alpha \beta}(X,t') \rangle_b \rho_d^\beta(\bar{X}^{\alpha \beta}_{0 \alpha \beta,t'}, t-t') \nonumber \\
   & & \quad + \sum_{\nu} \langle M_{\alpha \nu}^{\nu \alpha}(X,t') \rangle_b \rho_d^\alpha (\bar{X}^{\nu \alpha}_{0\alpha \nu , t'}, t-t') \Big) \;. \nonumber
\end{eqnarray}
In this expression the memory function appears in the form of an average over a bath distribution function conditional on the subsystem configuration space variables, $\langle M_{\alpha \beta}^{\alpha \beta}(X,t') \rangle_b \equiv \int dX_b M_{\alpha \beta}^{\alpha \beta}(X,t') \rho_b^{c}(X_b; R_0)$.  Since the memory function $M_{\alpha \beta}^{\alpha \beta}(X,t')$ involves evolution on the mean of the $\alpha$ and $\beta$ adiabatic surfaces, the average over bath initial conditions will result in an ensemble of such trajectories, each carrying an associated phase factor.  As a result, the bath ensemble average $\langle M_{\alpha \beta}^{\alpha \beta}(X,t') \rangle_b$ will decay on a time scale characterized by the decoherence time, $\tau_{decoh}$.  If the decoherence time is short compared to the decay of the populations we can make a Markovian approximation,
\begin{eqnarray}
   \langle M_{\alpha \beta}^{\alpha \beta}(X,t') \rangle_b &\approx& 2 \left(\int_0^\infty dt' \langle M_{\alpha \beta}^{\alpha \beta}(X,t') \rangle_b\right) \delta(t') \nonumber \\
   &\equiv& 2 m_{\alpha \beta}(X_0) \delta(t')\;.\label{eq:mark}
\end{eqnarray}
Applying this Markovian approximation to Eq.~(\ref{penult-PRME-folded}), we obtain,
\begin{eqnarray}
   & & \frac{\partial}{\partial t} \rho_s^\alpha (X_0,t) = \label{final-PRME-folded} \\
   & & \quad -\int dX_b i L_\alpha  e^{-i\mathcal{Q}L_\alpha t} \mathcal{Q}\rho_d^\alpha (X, 0) - \langle i L_\alpha \rangle_b \rho_s^\alpha(X_0,t) \nonumber \\
   & & \quad - \int_0^t dt' \langle i  L_\alpha e^{- i \mathcal{Q} L_\alpha t'} i \mathcal{Q} L_\alpha \rangle_b \rho_s^\alpha (X_0, t - t') \nonumber \\
   & & + \sum_\beta m_{\alpha \beta}(X_0)  j_{\alpha \to \beta} \rho_s^\beta (X_0, t) -  m_{\alpha \alpha} (X_0) \rho_s^\alpha (X_0, t) \;, \nonumber 
\end{eqnarray}
where
\begin{eqnarray}
   m_{\alpha \alpha}(X_0) & = & - \sum_\nu \int_0^\infty dt' \langle M_{\alpha \nu}^{\nu \alpha} (X, t') \rangle_b \;. \label{m-aa}
\end{eqnarray}
The use of the Markovian approximation results in the instantaneous action of two momentum shift operators on the density.  Consequently, the penultimate term in Eq.~(\ref{final-PRME-folded}) was rewritten to incorporate a single momentum shift operator whose action is $j_{\alpha \to \beta} f(X_0) = f(\bar{X}_{0 \alpha \beta}^{\alpha \beta})$
resulting from a transition from one single adiabatic surface to another single adiabatic surface:
\begin{eqnarray}
   j_{\alpha \to \beta} f(P_0) & = & j_{\alpha \beta}(X_0)j_{\alpha \beta}(X_0) f(P_0) \nonumber \\
   & = & f(P_0+\Delta {P_0}^{\alpha \beta}_{\alpha \beta})\;, \label{sing_mom_shift} \\
   \Delta {P_0}^{\alpha \beta}_{\alpha \beta} & = & \hat{d}_{\alpha \beta} \Big({\rm sgn}(P_0 \cdot \hat{d}_{\alpha \beta}) \label{eq:delta2P} \\
   & & \times \sqrt{( P_0 \cdot \hat{d}_{\alpha \beta})^2 + 2 \Delta E_{\alpha \beta}M_0} - (P_0 \cdot \hat{d}_{\alpha \beta})\Big) .  \nonumber
\end{eqnarray}
This momentum shift differs from that defined earlier by a factor of 2 in front of the energy difference since this operator captures the action of two jumps.  In the last term in Eq.~(\ref{final-PRME-folded}) the net effect of two momentum shift operators with reversed indices acting simultaneously is $j_{\nu \alpha}(\bar{X}_{0 \alpha \nu}) j_{\alpha \nu}(X_0) f(X)= f(X)$. Since these momentum shift operators are inverses of each other, there is no net shift.

As discussed in Appendix \ref{appendix:A}, the transition rate $M_{\alpha \beta}^{\alpha \beta}(X,t)$ captures the effect of two momentum shift operators.  Each of these operators imposes a condition on the subsystem kinetic energy that ensures transitions to and from the mean surface are allowed.  Consequently, the transition rate $m_{\alpha \beta}(X_0)$ inherits these conditions.  For example, if $\alpha < \beta$ then $m_{\alpha \beta}(X_0)$ is non-zero only if $(P_0 \cdot \hat{d}_{\alpha \beta})^2 / 2 M_0 > \Delta E_{\beta \alpha}$.  Conversely, if $\alpha > \beta$ there is no such restriction.  In contrast, the transition rate $m_{\alpha \alpha}(X_0)$ is non-zero only if $(P_0 \cdot \hat{d}_{\alpha \nu})^2 / 2 M_0 > \Delta E_{\nu \alpha} / 2$.  This condition arises from the fact that this contribution has its origin from transitions to the mean surface and then back to the original surface.    

\subsection*{Lift to full phase space}

Equation~(\ref{final-PRME-folded}) is still rather difficult to solve since it contains a convolution involving bath projected dynamics.  Often, the non-Markovian character of an equation can be removed by expanding the space upon which the equation is defined.  In the analysis above, the non-Markovian character arose by projecting out the bath variables to obtain a description in the subsystem phase space.  Consequently, by lifting this equation back into the full phase space we can recover the Markovian nature of the dynamics.  In the full phase space the equation of motion is given by
\begin{eqnarray}
   & & \frac{\partial}{\partial t} \rho_d^\alpha (X,t) = - i L_\alpha \rho_d^\alpha(X,t) \label{eq:RME} \\
   & &  + \sum_\beta  m_{\alpha \beta}(X_0)  j_{\alpha \to \beta}\rho_d^\beta (X, t) - m_{\alpha \alpha}(X_0) \rho_d^\alpha (X, t)  \;. \nonumber
\end{eqnarray}
It is easily verified that applying the projection operator algebra to this equation, using the projection operator defined in Eq.~(\ref{p-oper}), one obtains the subsystem evolution equation~(\ref{final-PRME-folded}).  Thus, when computing average values like those in Eq.~(\ref{reduced-density}), or their correlation function analogs discussed below, the master equation lifted to full phase space yields results identical to those of the non-Markovian equation~(\ref{final-PRME-folded}).

Through this analysis we have succeeded in finding a master equation description of the dynamics in the full phase space which incorporates the effects of decoherence.  The first term in Eq.~(\ref{eq:RME}) yields dynamics on single adiabatic surfaces.  The other terms correspond to contributions to the evolution due to nonadiabatic transitions between adiabatic states.  The nonadiabatic transition rates in these terms incorporate the effects of decoherence.  Given the nature of the dynamics generated by this master equation, there is a close connection to many currently-used surface-hopping schemes which will be discussed below.

\begin{figure}[tbp]
   \fbox{ \subfigure{
             \label{traj:1}  %
             \centering
                \includegraphics[width=8cm]{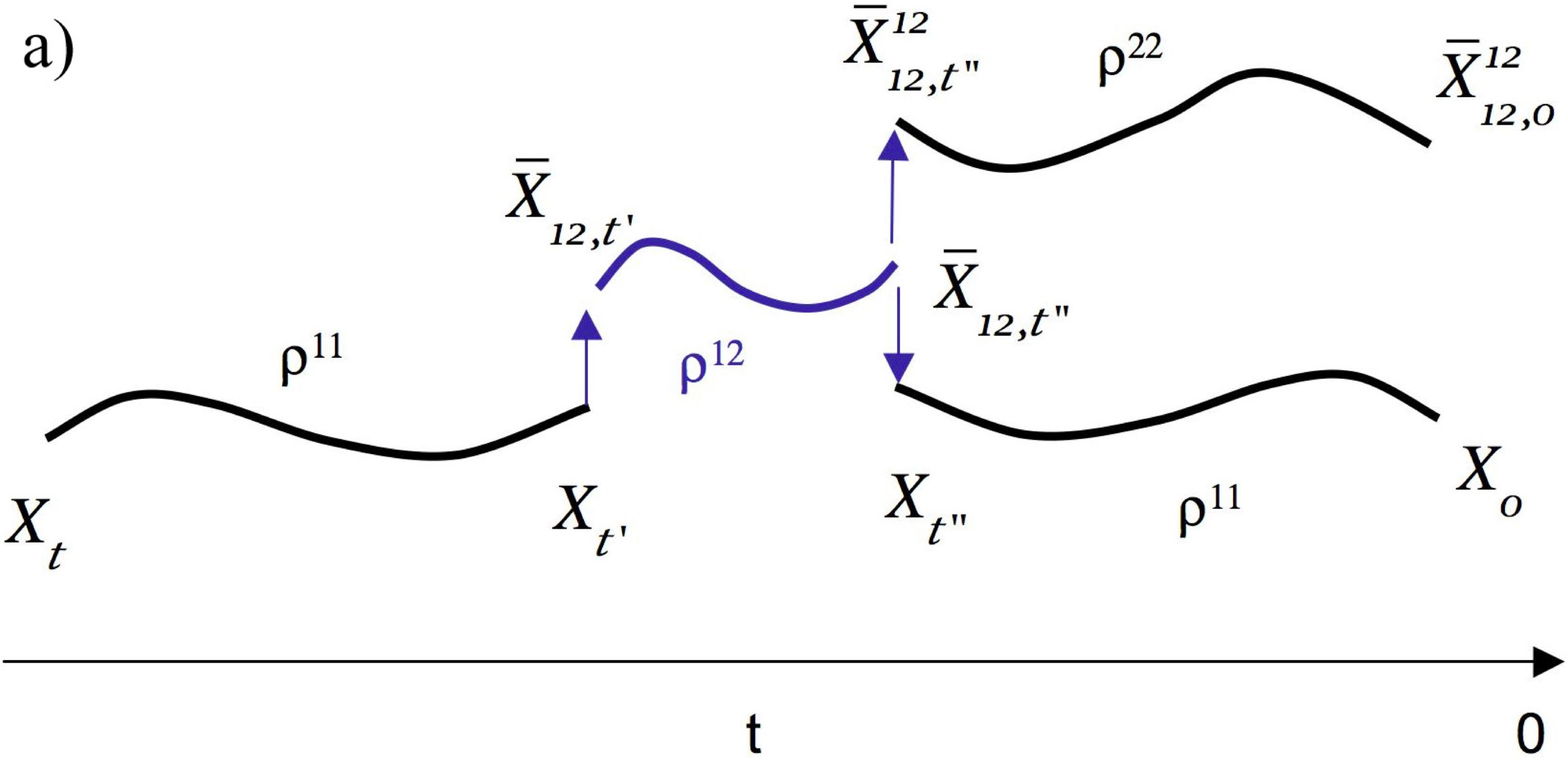}}
   }
   \fbox{ \subfigure{
             \label{traj:2}  %
             \centering
               \includegraphics[width=8cm]{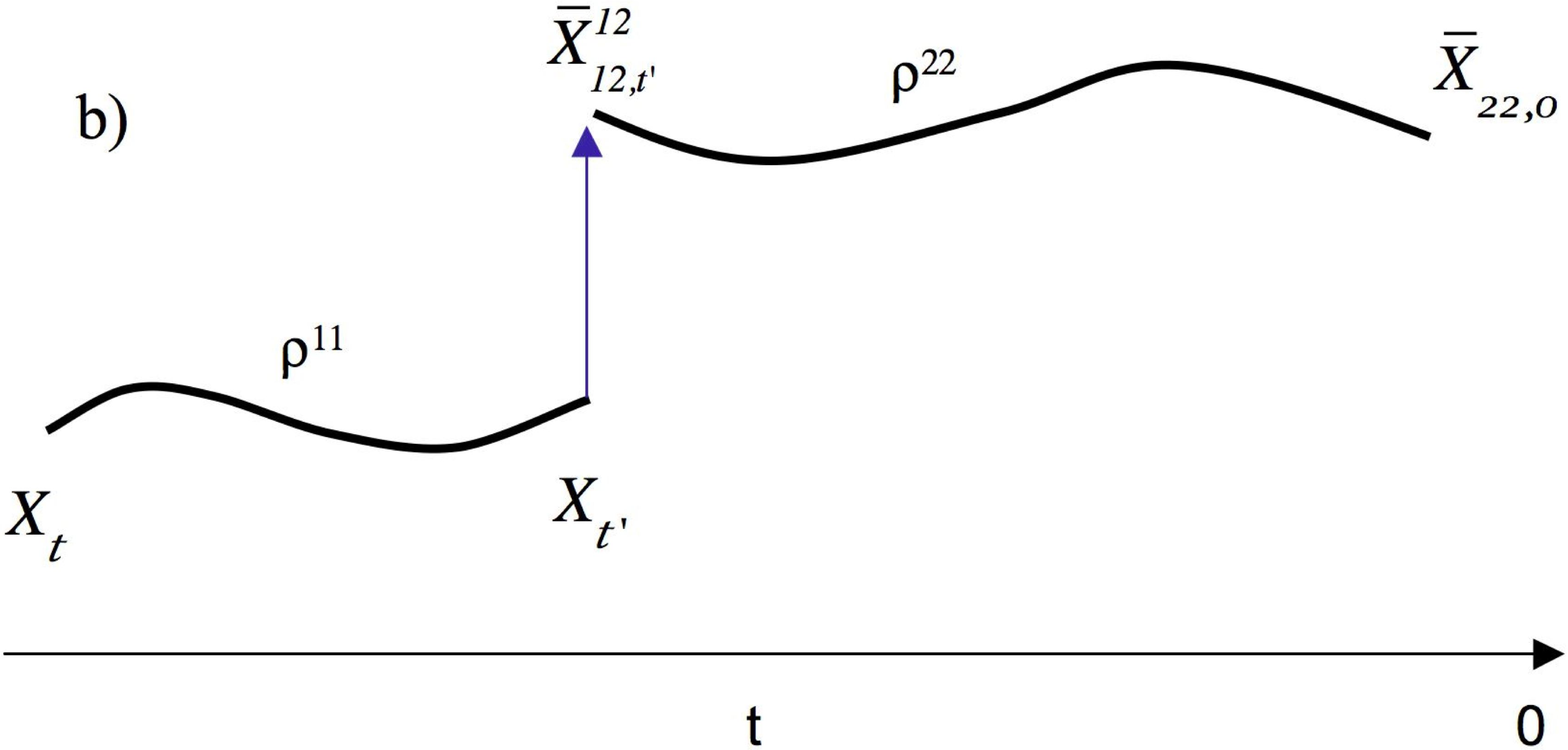}}
   }
   \caption{Trajectories that enter the solution of the quantum-classical Liouville
   and master equations.  \subref{traj:1} In quantum-classical Liouville dynamics, the system makes
   transitions between single adiabatic surfaces and coherently coupled states involving evolution on
   mean surfaces.  \subref{traj:2} In master equation evolution, the dynamics is restricted to
   single adiabatic surfaces.  All off-diagonal evolution is accounted for by the memory
   kernel.}
  \label{traj} %
\end{figure}

It is instructive to compare master equation and quantum-classical Liouville  dynamics.  The master equation~(\ref{eq:RME}), like the full quantum-classical Liouville equation~(\ref{Q-C Liouville}), can be simulated by following an ensemble of surface-hopping trajectories. The trajectories that enter in each description are shown in Fig.~\ref{traj}. We see that in full quantum-classical Liouville dynamics the system makes transitions between single adiabatic surfaces via coherently coupled off-diagonal states.  Coherence is created when such an off-diagonal state is entered and is destroyed when it is left. The average over the ensemble accounts for net destruction of coherence in the system as it evolves.  In contrast, the master equation evolves the classical degrees of freedom exclusively on single adiabatic surfaces with instantaneous hops between them.  Transitions from a diagonal state to a coherently coupled state and then back to the diagonal state, which play an important role in quantum-classical Liouville dynamics, are accounted for explicitly in master equation dynamics by $m_{\alpha \alpha} (X_0)$.  Each single (fictitious) trajectory accounts for an ensemble of trajectories that correspond to different bath initial conditions. In this connection the evolution in off-diagonal space is crucial: for a given initial subsystem coordinate, the choice of different bath coordinates will result in different trajectories on the mean surface.  Thus, it is the average over this collection of classical evolution segments that results in decoherence.  Consequently, this master equation in full phase space provides a description in terms of fictitious trajectories, each of which accounts for decoherence.  When the approximations that lead to the master equation are valid, this provides a useful simulation tool since no oscillatory phase factors appear in the trajectory evolution.

\section{Application to Reaction Rates} \label{sec:res}

In this section we apply the formalism developed above to calculate the rate constants of a reaction $A \rightleftharpoons B$.  For this reaction, the quantum-classical forward rate constant was derived earlier and is given by~\cite{kim052}
\begin{eqnarray}
   k_{AB}(t) & = &\frac{1}{n_A^{eq}} \sum_{\alpha} \sum_{\alpha' \geqslant \alpha}(2 - \delta_{\alpha' \alpha}) \label{rateconst} \\
   & & \quad \times \int dX \mbox{Re} \Big[N_B^{\alpha \alpha'} (X, t) W_A^{\alpha' \alpha} (X, \frac{i \hbar \beta}{2}) \Big], \nonumber
\end{eqnarray}
where $N_B^{\alpha \alpha'} (X, t)$ is the time evolved matrix element of the number operator for the product state B.  At $t=0$ this operator is diagonal in the adiabatic basis and its value, in this context, depends only on the subsystem coordinates $R_0$.  The spectral density function, $W_A^{\alpha' \alpha}(X, i \hbar \beta / 2)$,  accounts for the quantum equilibrium structure of the entire system.~\cite{sergi04, kim052}  The spectral density can be approximated by the form $W_A^{\alpha' \alpha}(X, i \hbar \beta / 2) \approx W_A^{\alpha' \alpha} (X_0, i\hbar \beta/2) \rho_b^c(X_b;R_0)$, such that it is factorized into subsystem and bath components.~\cite{kim05}  Performing the integration over the bath variables, the rate constant expression may be written as,
\begin{eqnarray}
   k_{AB}(t) & = & \frac{1}{n_A^{eq}} \int dX_0 dX_b N_B^{\alpha \alpha'}  (X, t) \rho_b^c (X_b; R_0) \nonumber \\
   & & \hspace{3.5cm}\times W_A^{\alpha' \alpha} (X_0, \frac{i \hbar \beta}{2}) \label{rate_bath_average} \\
   & = & \frac{1}{n_A^{eq}} \int dX_0 \mbox{Re} \left[\langle N_B^{\alpha \alpha'}  (X, t) \rangle_b W_A^{\alpha' \alpha} (X_0, \frac{i \hbar \beta}{2}) \right]. \nonumber
\end{eqnarray}
We see that the calculation of the rate coefficient entails knowledge of the bath average of the time-evolved species variable and sampling from the subsystem spectral density function.  The time evolution of this species variable may be calculated using mixed quantum-classical dynamics.~\cite{kim062}  In general, the subsystem spectral density contains both diagonal and off-diagonal components; therefore, both diagonal and off-diagonal components of the species operator contribute to the computation of the rate coefficient.  Previous work has shown that the off-diagonal contributions to the rate constant are negligible,~\cite{kim06} allowing one to consider only diagonal contributions.

The computation of the time evolution of the bath averaged species variable is completely analogous to the calculation of the subsystem density matrix leading to Eq.~(\ref{final-PRME-folded}); however, now the analysis must be carried out starting with the quantum-classical Heisenberg equation of motion,~\cite{kapral99}
\begin{equation}
   \frac{d }{d t}A^{\alpha \alpha'}(X, t) = \sum_{\beta \beta'} i \mathcal{L}_{\alpha \alpha', \beta \beta'} A^{\beta \beta'}(X,   t). \label{Q-C Liouville-Heisenberg}
\end{equation}
The rate coefficient can be computed from the expression in the first line of Eq.~(\ref{rate_bath_average}) using the lifted form of the evolution equation for the diagonal elements of a dynamical variable,
\begin{eqnarray}
   & &\frac{d}{d t} A^{\alpha \alpha} (X, t) = iL_\alpha (X_0) A^{\alpha \alpha} (X, t) \label{eq:RME_var} \\
      & & \qquad + \sum_{\beta} m^\dag_{\alpha \beta}(X_0) j_{\alpha \to \beta}A^{\beta \beta} (X, t) - m^\dag_{\alpha \alpha}(X_0) A^{\alpha \alpha} (X, t) \;, \nonumber
\end{eqnarray}
where the memory function, $m^\dag$, is the adjoint of $m$ defined previously in Eq.~(\ref{eq:mark}).  The effects of decoherence that lead to this expression restrict the evolution of the observable to its diagonal components.  Therefore, one only needs to consider the diagonal terms of the subsystem spectral density in the calculation of the rate coefficient.

\subsection*{Model system}\label{model}

As an application of this formalism, we consider a simple model for a quantum rate process that has been studied earlier using mixed quantum-classical dynamics.~\cite{kim06}  The investigation of this model allows us to assess the validity of the Markovian approximation, Eq.~(\ref{eq:mark}), and the utility of the master equation for calculating the rate coefficient

The model is a two-level system to which we couple $\nu$ oscillators.  The \textit{subsystem} consists of the two-level quantum system bilinearly coupled to a non-linear oscillator with phase space coordinates $(R_0, P_0)$ governed by a symmetric quartic potential, $V_q (R_0) = a R_0^4/4 - b R_0^2/2$.  The \textit{bath} consists of $\nu - 1 = 300$ harmonic oscillators whose frequencies $\omega_j$, are distributed with Ohmic spectral density that depends on $\xi_K$, the Kondo parameter.~\cite{makri99}  The bath is bilinearly coupled to the subsystem oscillator such that the quantum system does not directly interact with the bath; it only feels its effects through the coupling to the quartic oscillator.  As discussed elsewhere, it has been argued that this model captures many of the essential features of condensed phase proton transfer processes.~\cite{hanna05, kim06}

\begin{figure} 
    \includegraphics[width=8cm]{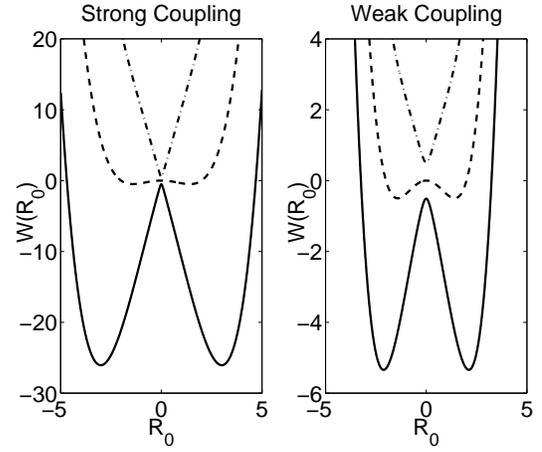}
    \caption{Plots of free energy vs $R_0$ for strong and weak coupling cases.  The parameters are: $\gamma_0 = 10.56$ for strong coupling and $\gamma_0 = 2.64$ for weak coupling between the two-level quantum system and the quartic oscillator.  The other parameters are the same for both cases: $\Omega=0.51$, $\beta = 0.5$, $\xi_K = 2$, $A=0.5$ and $B=1$.  These parameters were chosen to give a well-defined rate process with a significant number of re-crossing events.  The small energy gap ensures that the majority of trajectories satisfy the energetic requirements for nonadiabatic transitions given by (\ref{eq:deltaP}) and (\ref{eq:delta2P}), and the parameter $\beta$ was chosen to be small enough to satisfy the high temperature approximation.  All other parameters in the Ohmic spectral density are the same as those used in earlier studies,~\cite{kim06} and the results are presented in the same dimensionless units as those used in previous studies.~\cite{sergi03}.}
    \label{FE}
\end{figure}

Using a diabatic representation, the Hamiltonian for this system is,
\begin{eqnarray}
   & &\textbf{H} = \left(\begin{array}{cc}V_q(R_0) + \hbar \gamma_0 R_0  & -\hbar \Omega \\ -\hbar \Omega & V_q(R_0) - \hbar \gamma_0 R_0\end{array}\right) \label{hamiltonian} \\
   & & + \left(\frac{P_0^2}{2M_0} + \sum_{j=1}^{\nu - 1} \frac{P_j^2}{2M_j} + \frac{M_j \omega_j^2}{2} \left( R_j - \frac{c_j}{M_j \omega_j^2} R_0 \right)^2 \right) \mathbf{I}. \nonumber
\end{eqnarray}
The solution of the eigenvalue problem for this Hamiltonian yields the adiabatic eigenstates, $|\alpha;R_0 \rangle$, and eigenvalues $E_\alpha (R) = V_q(R_0) + V_b(R_b;R_0) \mp \hbar \sqrt{\Omega^2 + (\gamma_0 R_0)^2}$, where $2 \Omega$ is the adiabatic energy gap.  The adiabatic free energy surfaces, $W_\alpha(R_0) = V_q(R_0) \mp \hbar \sqrt{\Omega^2 + (\gamma_0 R_0)^2}$ are sketched in Fig.~\ref{FE}.  In this figure we also show the mean free energy surface, $W_{12}(R_0) = (W_1 + W_2)/2 = V_q(R_0)$, which plays an essential role in the calculation of the memory function.

The simulations of quantum-classical Liouville dynamics were carried out using the sequential short-time propagation algorithm~\cite{kapral&mackernan&ciccotti02} in conjunction with the momentum-jump approximation~\cite{kapral&ciccotti01, hanna05} and a bound on the observable.~\cite{hanna05}  The initial positions and momenta of the quartic oscillator and bath were sampled from the classical canonical density function.  The details of these methods can be found elsewhere.~\cite{hanna05, kim06, kapral&mackernan&ciccotti02}  The simulations of the master equation consist of two parts which we describe below.  First we compute $m_{\alpha \beta} (X_0)$ in an independent calculation involving evolution on the mean surface.  Then we use this result in the sequential short-time propagation algorithm restricted to single adiabatic surfaces.

\subsection*{Calculation of $m_{\alpha \beta} (X_0)$}

\begin{figure} 
      \includegraphics[width=8cm]{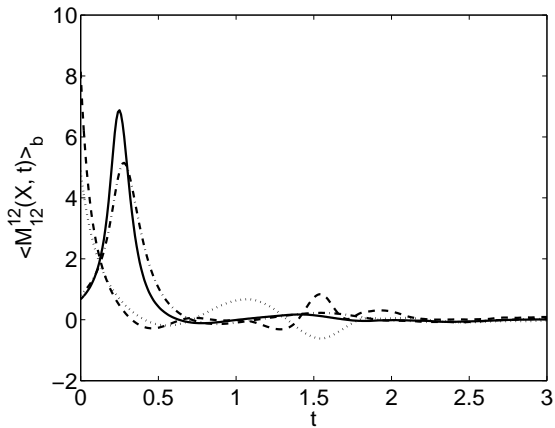}
   \caption{Plot of the bath averaged memory function $\langle M_{12}^{12}(X, t) \rangle_b$ versus time for $\gamma_0 = 2.64$, solid line: $R_0 = -0.55$, $P_0 = 3.2$, dotted line: $R_0 = 0.4$, $P_0=2.4$, dashed line: $R_0 = -0.25$, $P_0 = -3.4$, dot-dash line: $R_0 = 0.6$, $P_0 = -2.2$.  Here we see that for a range of choices of $X_0$, this function decays quickly.}
   \label{avg_mk}
\end{figure}

In order to investigate the validity of the Markovian approximation we calculate $\langle M_{\alpha \beta}^{\alpha \beta}(X, t)\rangle_b$, as a function of time.  From Eq.~(\ref{rate_bath_average}), this average is weighted by $\rho_b^c(X_b; R_0)$, the Wigner representation of the quantum bath distribution conditional on the subsystem coordinate.  In general the determination of the quantum distribution function is a difficult problem; however, it is known for a harmonic bath,~\cite{wignermethod67} and may be used to account for quantum bath effects.  In the Conclusions we comment briefly on quantum bath effects in our formalism.  In our calculations we use the high temperature limit where the classical canonical equilibrium density, conditional on the subsystem configuration, provides a good approximation.~\cite{sergi04}  

\begin{figure} 
      \includegraphics[width=8cm]{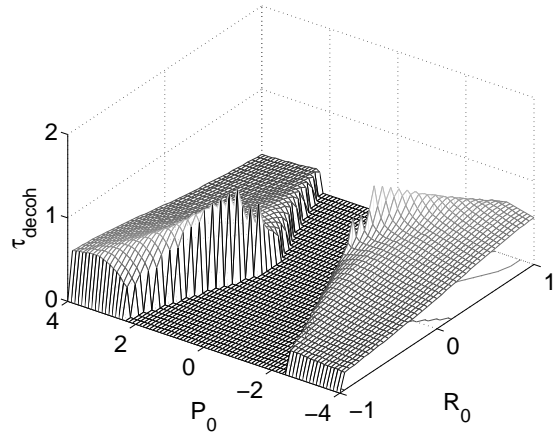}
   \caption{Plot of $\tau_{decoh}$ corresponding to upward transitions $1 \to 2$ vs $R_0$ and $P_0$ for $\gamma_0 = 2.64$.}
      \label{tdec}
\end{figure}

The quantity $\langle M_{\alpha \beta}^{\alpha \beta}(X, t)\rangle_b$ involves the product of the initial value of $D_{\alpha \beta}$, the phase factor $\mathcal{W}_{\alpha \beta}$, and $D_{\alpha \beta}$ at a time-evolved phase point. The latter two quantities may be obtained from adiabatic dynamics on the mean surface for a given $X_0$. The bath averaged memory function, $\langle M_{\alpha \beta}^{\alpha \beta} (X, t) \rangle_b$ may be computed from an average over an ensemble of trajectories, each with a fixed initial value of $X_0$ and bath coordinates drawn from the phase space distribution $\rho_b^c(X_b;R_0)$.  As discussed above, the bath average of this oscillatory function provides a mechanism for its decay, characterized by the decoherence time, $\tau_{decoh}$. This time will depend on the subsystem coordinate $X_0$.  In Fig.~\ref{avg_mk} we plot $\langle M_{\alpha \beta}^{\alpha \beta}(X, t) \rangle_b$ as a function of time for several subsystem coordinate values and show that the bath averaged memory function does indeed decay on a rapid time scale.  Figure~\ref{tdec} shows how the decoherence time, taken as the first zero crossing of $\langle M_{\alpha \beta}^{\alpha \beta}(X, t) \rangle_b$, depends on the phase space coordinate $X_0$.  In the allowed phase space regions, the decoherence time is a relatively weak function of the phase space coordinates, with the exception of some localized regions where it varies strongly.  From these results we may compute the mean decoherence time and find $\tau_{decoh} = 0.41 \pm 0.09$ (weak coupling) and $\tau_{decoh} = 0.17 \pm 0.02$ (strong coupling).   In order for the Markovian approximation to be valid, the decoherence time must be short compared to the characteristic decay times of the correlation function that determines the rate constant.

\begin{figure} 
      \subfigure{
             \label{msurf:1}  %
             \centering
                 \includegraphics[width=8cm]{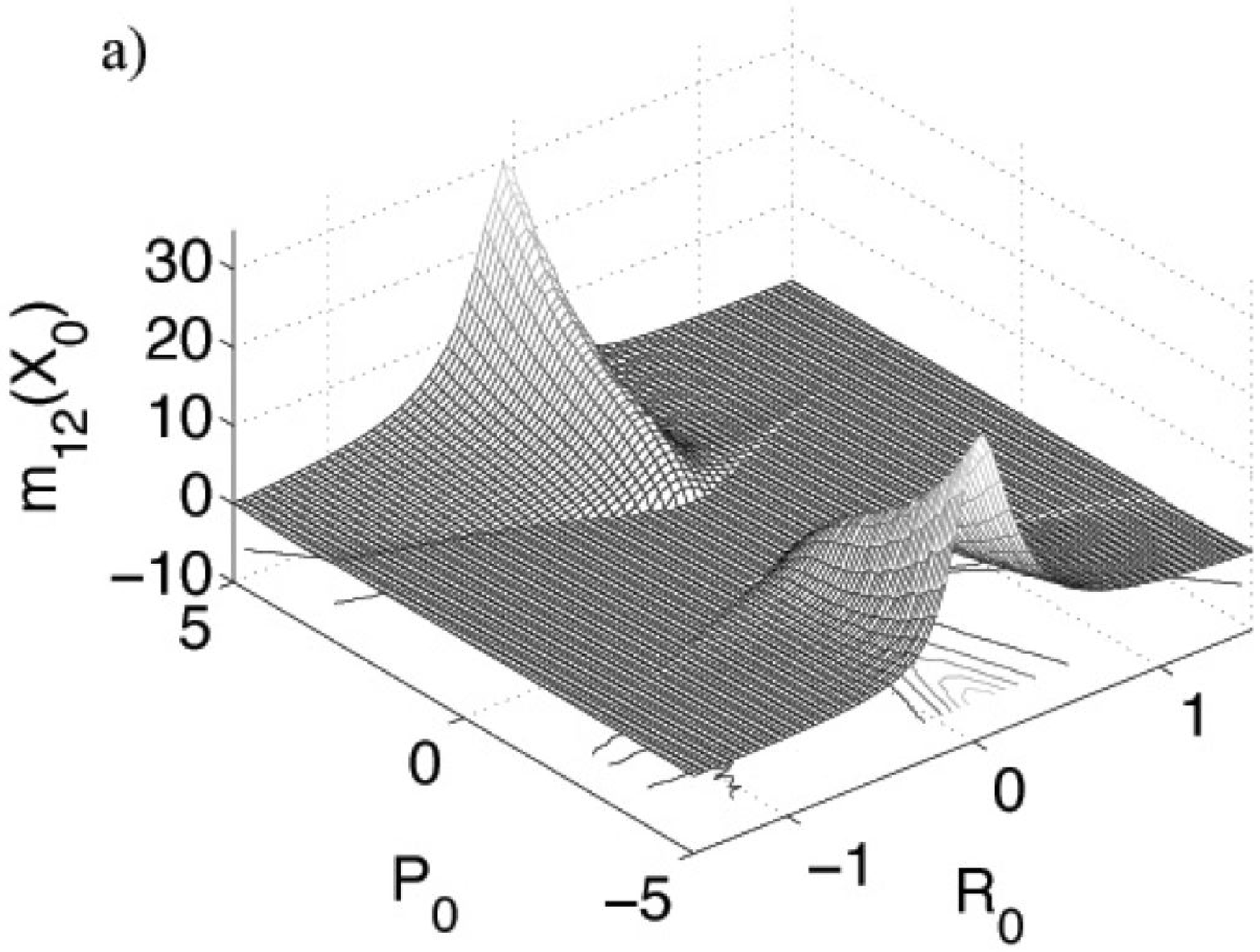}
             }
   \subfigure{
             \label{msurf:2}  %
             \centering
                \includegraphics[width=8cm]{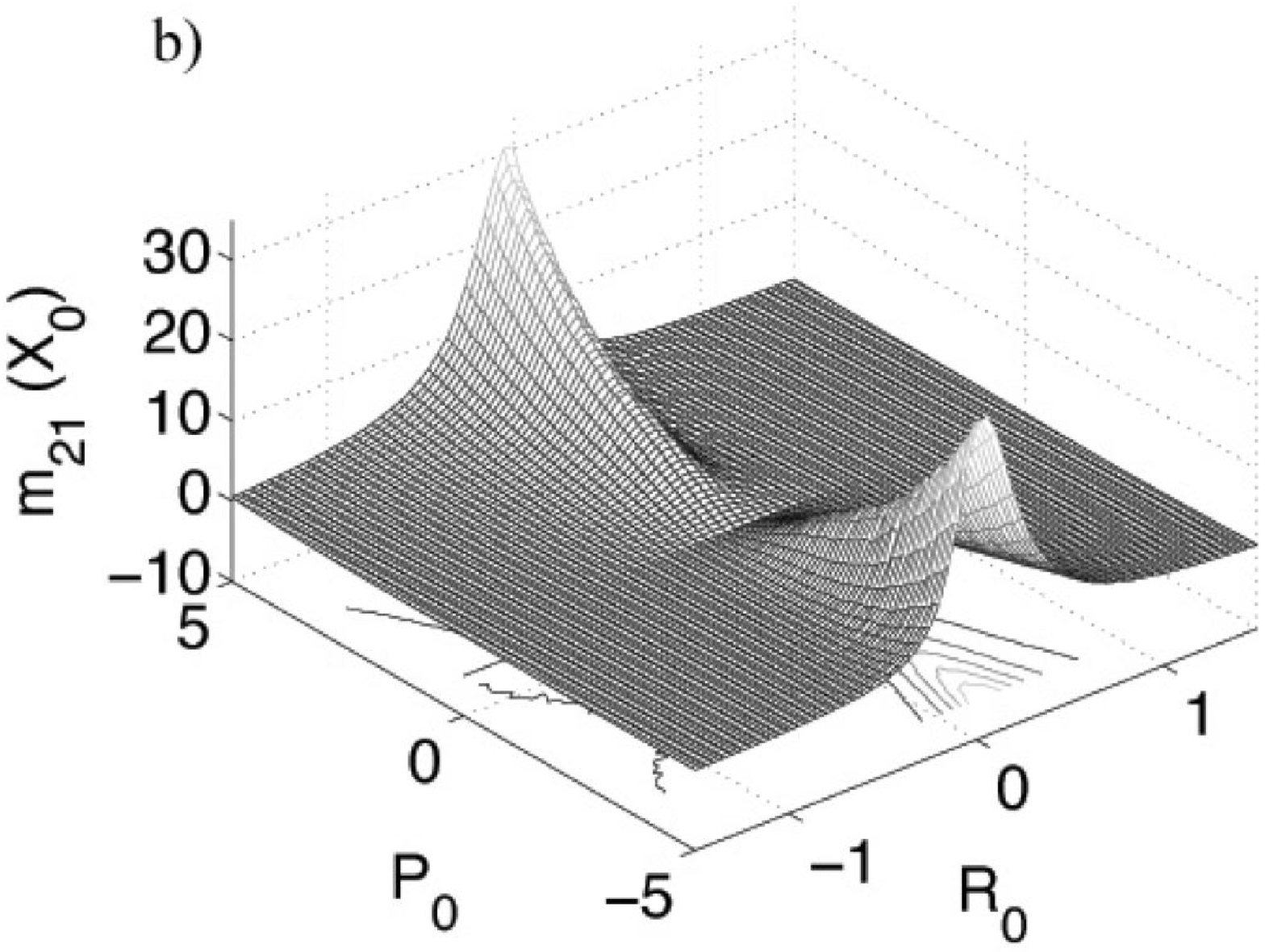}
             }
                \caption{Plots of $m_{\alpha \beta} (R_0, P_0)$ versus $R_0$ and $P_0$ for $\gamma_0 = 2.64$.  \subref{msurf:1} $m_{1 2} (R_0, P_0)$ portions of the surface have a value of zero, corresponding to regions where transitions are forbidden due to insufficient kinetic energy in the subsystem.  \subref{msurf:2} $m_{2 1}(R_0, P_0)$ does not have this feature as it corresponds to downward transitions where the subsystem gains kinetic energy.}\label{msurf} %
\end{figure}

Simulation of the master equation requires knowledge of the transition rates $m_{\alpha \beta} (X_0) $.  These quantities were obtained by numerically integrating the time dependent memory function discussed above.  In this calculation one must ensure that for a given $X_0$ the transition is allowed.  Otherwise $m_{\alpha \beta} (X_0)$ is assigned a value of zero for that choice of subsystem coordinates.  These restrictions were discussed in Sec.~\ref{sec:RME}. This process is repeated for a range of $X_0$ values generating the surface, $m_{\alpha \beta} (R_0, P_0)$.  We obtain a different surface for each transition (see Fig.~\ref{msurf}).  

The structure of these transition-rate surfaces is due entirely to classical evolution of $X_0$ along the mean surface.  It is precisely this evolution that leads to spread in the ensemble of trajectories giving rise to decoherence.  Thus, even though the evolution we are ultimately interested in calculating is entirely in diagonal space, the probability of the nonadiabatic transitions is calculated from the off-diagonal or coherent evolution segments dependent on $X_0$.  In this way decoherence is accounted for in the formalism.

\subsection*{Simulation of the master equation}

Once the surfaces, $m_{\alpha \beta} (X_0)$ are obtained, Eq.~(\ref{eq:RME_var}) is simulated using the sequential short-time propagation algorithm where the probabilities of nonadiabatic transitions are given by $\Pi = |m_{\alpha \beta}|\Delta t/(1 + |m_{\alpha \beta}|\Delta t)$.~\cite{kapral&mackernan&ciccotti02}  Note that the value of $\Pi$ is determined at each time step using the value of $m_{\alpha \beta}(X_0)$ corresponding to the specific value of $X_0$ at that time.  The initial sampling is taken from the spectral density function where the bath distribution is given by the conditional density, $\rho_b^c(X_b; R_0)$.

\begin{figure}
      \subfigure{
             \label{result:1}  %
             \centering
                 \includegraphics[width=8cm]{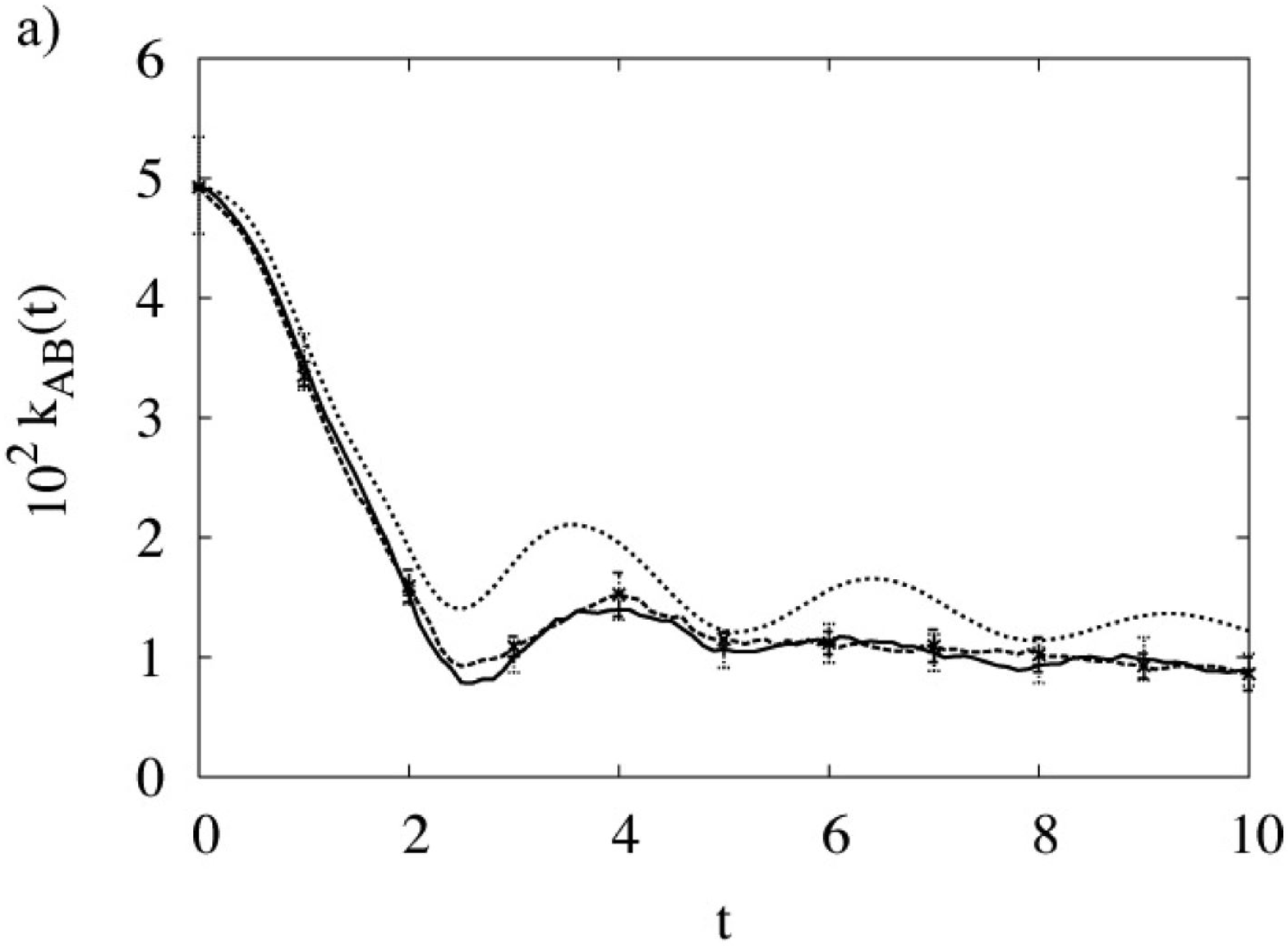}
             }
   \subfigure{
             \label{result:2}  %
             \centering
                \includegraphics[width=8cm]{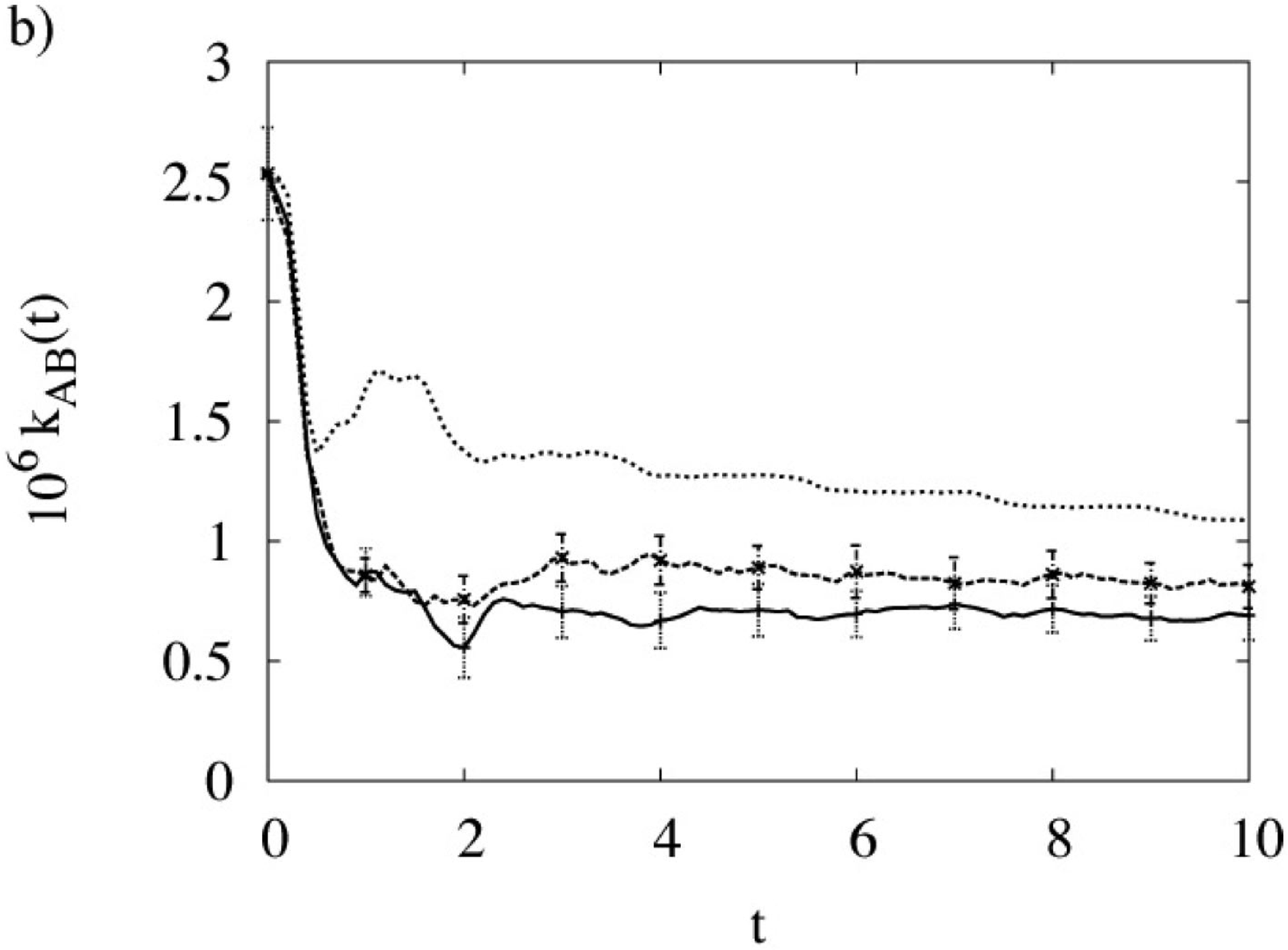}
             }
                \caption{Plots of the time dependent rate coefficient as a function of time.  \subref{result:1} $\gamma_0 = 2.64$ (weak coupling) and \subref{result:2} $\gamma_0 = 10.56$ (strong coupling).  In these plots the upper dotted curve is from adiabatic dynamics, the middle curve is from master equation dynamics, and the lowest solid curve is from quantum-classical Liouville dynamics.} 
\label{result} %
\end{figure}

The results of the calculation of the forward time dependent rate coefficient, $k_{AB}(t)$,  are shown in Fig.~\ref{result}.  The figure compares the rate coefficients using adiabatic, master equation, and quantum-classical Liouville dynamics.  As expected the plots show rapid decay on a time scale $\tau_{mic}$ to a plateau region characterized by a much slower decay on the macroscopic chemical relaxation time scale, $\tau_{chem} \approx 67$ for weak coupling and $\approx 1.3 \times 10^6$ for strong coupling.~\cite{kapral&consta&mcwhirter98}

In Fig.~\ref{result}, we see that the short-time decay portion of the rate coefficient given by the master equation simulation is in agreement with the quantum-classical result.  The time scale of this decay, $\tau_{mic}\approx 4$ in the weak coupling case and $\approx 2.5$ for strong coupling, is about one order of magnitude larger than the average decoherence time $\tau_{decoh} \approx 0.41$ for weak coupling, and $\approx 0.17$ for strong coupling as discussed above.  From the figures we conclude that indeed $\tau_{decoh} \ll \tau_{mic} \ll \tau_{chem}$.  This inequality provides the conditions for the applicability of the Markovian approximation used to derive the master equation. The plateau regions for both quantum-classical Liouvile and master equation dynamics have lower values than those for adiabatic dynamics. The smaller rate constant for nonadiabatic dynamics is due to enhanced barrier recrossing as a result of motion on either the excited state or mean surfaces. The plateau value using master equation dynamics in the strong coupling case is slightly higher than that obtained using quantum-classical Liouville dynamics. This likely arises from the fact that in quantum-classical Liouville dynamics the system evolves on the mean surface for long times, allowing trajectories to re-enter the region of high nonadiabatic coupling where quantum transitions take place.  Thus, the rate coefficient is reduced due to recrossings in the barrier region. In general, for both weak and strong coupling, the master equation provides quite a good description of the rate coefficient data.

\section{Conclusion}

The master equation calculations presented above bear many similarities to surface-hopping schemes that have been used previously to simulate nonadiabatic dynamics of quantum-classical systems~\cite{tully90, tully91, hammesschiffer94, coker95, rossky&prezhdo97, rossky&bittner95}. It is useful to comment on some of the similarities and highlight the differences.  In most surface-hopping schemes, and in our master equation dynamics, the classical degrees of freedom evolve on single adiabatic potential energy surface segments according to Newton's equations of motion governed by Hellmann-Feynman forces. This dynamics should be contrasted with the trajectory evolution in quantum-classical Liouville dynamics, where the trajectory segments of the classical degrees of freedom evolve on single adiabatic surfaces as well as mean surfaces. The differences between our master equation dynamics and other surface-hopping methods lie in the prescription for quantum transitions and the manner in which decoherence is incorporated into the theory.  For example, in the fewest-switches surface hopping scheme,~\cite{tully90, tully91, hammesschiffer94} the probability of a transition depends on the nonadiabatic coupling matrix elements and the off-diagonal elements of the density matrix.  In our master equation the probabilities of quantum transitions are determined by a Monte Carlo sampling based on the magnitudes of our phase space dependent transition rates, $m_{\alpha \beta}(X_0)$, and the sampling algorithm reweights averages so that no bias is introduced.  Decoherence is accounted for in the fewest-switches simulation by collapsing the density matrix onto a diagonal state depending on certain conditions such as motion outside a window of strong coupling.~\cite{hammesschiffer94} In our calculation decoherence effects have been incorporated into the calculation of the transition rates.

Decoherence has also been incorporated into the formulations of surface-hopping methods using other physical principles. The idea in such methods is to include the effects of decoherence in single trajectories, much like the description our master equation provides.  For example, in the methods developed by Rossky et. al.~\cite{rossky&prezhdo97,  rossky&bittner95, rossky&wong02} decoherence is introduced through an additional term in the evolution equation that accounts for the quantum dispersion about each classical phase space coordinate in the bath.  In this way each of the trajectories in the full phase space experiences decoherence.  The quantum dispersion of the bath is not included in the model calculations presented above but it is easily accounted for in our theory.  In our master equation decoherence arises through averaging over the bath phase space variables which were taken to be classically distributed for high temperatures.  The rate coefficient formalism in Eq.~(\ref{rate_bath_average}) involves sampling from the full quantum spectral density function thus incorporating quantum dispersion in the bath coordinates.  Such quantum effects have already been investigated in the context of quantum-classical Liouville dynamics~\cite{kim062} and for the temperatures used in our calculations, these effects are very small.  Regardless of whether the bath is treated classically or quantum mechanically, decoherence enters our master equation through the forms of the transition rates and not as an additional term in the equation of motion.   

Finally, we remark that the simulation scheme for master equation dynamics has a number of attractive features when compared to quantum-classical Liouville dynamics.  The solution of the master equation consists of two numerically simple parts.  The first is the computation of the memory function which involves adiabatic evolution along mean surfaces.  Once the transition rates are known as a function of the subsystem coordinates, the sequential short-time propagation algorithm may be used to evolve the observable or density.  Since the dynamics is restricted to single adiabatic surfaces, no phase factors enter the calculation increasing the stability of the algorithm.  For complex reaction coordinates which are arbitrary functions of the bath coordinates the calculation of the transition rates will be more difficult and time consuming.  Future research will determine if the master equation can be applied easily to realistic general many-body systems.  Nevertheless, the results reported in this paper have served to provide a basis for an understanding of the domain of validity of master equation approaches to quantum-classical nonadiabatic dynamics based on decoherence.

\section*{Acknowledgements}

This work was supported in part by a grant from the Natural Sciences and Engineering Research Council of Canada.

\begin{appendix}

\section{Reduction to Memory Function}\label{appendix:A}

Starting from the form of the memory kernel operator given in Eq.~(\ref{mem kern}), we may reduce this operator to a function. Without loss of generality, we assume that the adiabatic basis is real so that, $\mathcal{C}_{\alpha \beta} = \mathcal{C}^*_{\alpha \beta}$. Furthermore, taking the definition of $\mathcal{J}$, and acting with the operator $j_{\alpha \nu}(X_0)$ coming from the leftmost $\mathcal{J}^{d,o}$ operator on all operators to its right, the memory kernel operator may be written as
\begin{eqnarray}
   & & \mathcal{M}_{\alpha \beta} (t) \nonumber \\
   & & \qquad = \sum_{\nu \nu'} D_{\alpha \nu}(X_0) 2{\rm Re} \Big[ \mathcal{U}^o_{\alpha \nu, \beta \nu'}(\bar{X},t) + \mathcal{U}^o_{\alpha \nu, \nu' \beta}(\bar{X},t)\Big] \nonumber \\
   && \qquad \qquad \times D_{\nu' \beta}(\bar{X}_0) j_{\nu' \beta}(\bar{X}_0)j_{\alpha \nu}(X_0) \label{memory kernel4}\;.
\end{eqnarray}
In the above expression we have introduced the off-diagonal propagator, $\mathcal{U}^o_{\alpha \nu, \beta \nu'}(X, t)
=\left(e^{-i \mathcal{L}^o(X) t} \right)_{\alpha \nu, \beta \nu'}$.  In the case of a two level system this propagator is given exactly by
\begin{equation}
    \mathcal{U}^o_{\mu \mu', \nu \nu'} (t) = \mathcal{W}_{\mu \mu'}(t, 0) e^{-i L_{\mu \mu'} (X) t} \delta_{\mu \nu} \delta_{\mu' \nu'} \;, \label{2ls-prop}
\end{equation}
for $\mu, \mu', \nu, \nu' = 1, 2$ and $\mu \neq \mu', \nu \neq \nu'$.  Here we used the fact that~\cite{kapral99}
\begin{eqnarray}
   e^{(-i\omega_{\mu \mu'}-iL_{\mu \mu'}) (t-t')}& = & e^{-i\int_t^{t'} d\tau \; \omega_{\mu \mu'}(R_{0\mu \mu', \tau})} e^{-iL_{\mu \mu'} (t-t')} \nonumber \\
   &\equiv&  \mathcal {W}_{\mu \mu'}(t,t')e^{-iL_{\mu \mu'}(t-t')} \;, \label{eq:wphase}
\end{eqnarray}
to express the operator as a product of a phase factor and a classical propagator.  We note that the only off-diagonal propagator matrix elements that contribute to the dynamics here are $\mathcal{U}^o_{12, 12} = \mathcal{U}^{o*}_{21, 21}$.

Recall from the definition of the momentum shift operator, Eq.~(\ref{eq:deltaP}), that transitions can only occur if there is sufficient momentum in the subsystem to make a transition to or from a mean surface.  Otherwise the transitions are not allowed.  Using the above form of the off-diagonal propagator in Eq.~(\ref{memory kernel4}), the action of the memory kernel operator on some arbitrary function of the phase space variables, $f(X_0, X_b)$, takes the following form:
\begin{eqnarray}
   && \mathcal{M}_{\alpha \beta} (t) f(X_0, X_b) = \nonumber \\
   && \quad \qquad \delta_{\alpha \beta}\sum_{\nu} 2{\rm Re} \Big[ \mathcal{W}_{\alpha \nu}(t,0)\Big] D_{\alpha \nu}({X_0}_{\alpha \nu})e^{-iL_{\alpha \nu}(\bar{X}),t} \nonumber \\
   && \qquad \qquad \times    D_{\nu \alpha}(\bar{X}_{0 \alpha \nu})  j_{\nu \alpha}(\bar{X}_{0 \alpha \nu}) f(\bar{X}_{0 \alpha \nu}, X_b)\nonumber \\
   && \quad \qquad + 2{\rm Re} \Big[ \mathcal{W}_{\alpha \beta}(t,0)\Big] D_{\alpha \beta}({X_0}_{\alpha \beta}) e^{-iL_{\alpha \beta}(\bar{X})t} \label{memory-kernel-approx} \\
   && \qquad \qquad \times    D_{\alpha \beta}(\bar{X}_{0 \alpha \beta})  j_{\alpha \beta}(\bar{X}_{0 \alpha \beta}) f(\bar{X}_{0 \alpha \beta}, X_b) \;. \nonumber
\end{eqnarray}
The arguments of $f$ reflect the fact that we have acted with the rightmost momentum shift operator on the function.  If we now consider the action of the classical propagators that enter in the memory kernel operator we obtain,
\begin{eqnarray}
   &&e^{-iL_{\alpha \nu}(\bar{X}_{\alpha \nu})t} D_{\nu \alpha}(\bar{X}_{0 \alpha \nu})j_{\nu \alpha} (\bar{X}_{0 \alpha \nu}) f(\bar{X}_{0 \alpha \nu}, X_b)\nonumber \\
   && \qquad = D_{\nu \alpha}(\bar{X}_{0 \alpha \nu ,t}) j_{\nu \alpha}(\bar{X}_{0 \alpha \nu ,t}) f(\bar{X}_{0 \alpha \nu ,t}, X_{b, t})\nonumber \\
   && \qquad = D_{\nu \alpha}(\bar{X}_{0 \alpha \nu ,t}) f(\bar{X}^{\nu \alpha}_{\alpha \nu ,t}, X_{b,t})\;. \label{Jaction}
\end{eqnarray}
In the last line we denoted the indices coming from the action of the second momentum shift operator as superscripts ($j_{\nu \alpha}(\bar{X}_{0 \alpha \nu }) f(\bar{X}_{0 \alpha \nu}, X_b) = f(\bar{X}^{\nu \alpha}_{0 \alpha \nu }, X_b)$).  Substituting Eq.~(\ref{Jaction}) in the expression (\ref{memory-kernel-approx}) for the memory kernel we obtain,
\begin{eqnarray}
   & &\mathcal{M}_{\alpha \beta}(X,t) f(X_0, X_b) = \delta_{\alpha \beta} \sum_\nu M_{\alpha \nu}^{\nu \alpha}(X, t) f(\bar{X}^{\nu \alpha}_{\alpha \nu ,t}, X_{b,t}) \nonumber \\
   & & \qquad \quad + M_{\alpha \beta}^{\alpha \beta} (X, t) f(\bar{X}^{\alpha \beta}_{\alpha \beta ,t}, X_{b,t}) \;, \label{mem_parts}
\end{eqnarray}
where the definition of $M$ is given in Eq.~(\ref{mk_function}).

\section{Subsystem Master Equation}\label{appendix:B}

In this appendix we focus on the equation of motion for the subsystem density matrix. In order to simplify the notation in the following calculation, it is convenient to write the generalized master equation~(\ref{meqn-wc}) in a more formal and compact form. Letting,
\begin{eqnarray}
   & & \sum_\beta M_{\alpha \beta}^{\alpha \beta} (X, t') {\rho}_d^\alpha(\bar{X}_{0 \alpha \beta, t'}^{\alpha \beta}, X_{b, t'}, t-t') \nonumber\\
   & & \qquad + \sum_\nu M_{\alpha \nu}^{\nu \alpha} (X, t') {\rho}_d^\alpha(\bar{X}_{0 \alpha \nu, t'}^{\nu \alpha}, X_{b, t'}, t-t')  \nonumber \\
   & & \equiv \left( M(X, t') {\rho}_d (\bar{X}_{t'}, t - t') \right)_\alpha \;. \label{mem-simplifn}
\end{eqnarray}
we can write Eq.~(\ref{meqn-wc}) as
\begin{eqnarray}
   &&   \frac{\partial }{\partial t}\rho_d (X, t) = - i L_d \rho_d (X, t) \label{meqn-wc-formal} \nonumber\\
   & & \qquad \qquad + \int_0^t dt' M(X,t')\rho_d (\bar{X}_t, t-t') \;.
\end{eqnarray}

Starting from Eq.~(\ref{meqn-wc-formal}), we use standard projection operator methods~\cite{zwanzig61} to obtain the evolution equation for the subsystem density matrix.  If we let $\rho_b^{c}(X_b;R_0)$ be the bath equilibrium density matrix conditional on the configuration of the directly coupled $R_0$ subsystem coordinates, we may define the projection operator as in Eq.~(\ref{p-oper}) and it's complement, by $\mathcal{Q} = 1 - \mathcal{P}$. Note that $\mathcal{P} \rho_d(X,t)=\rho_b^{c}(X_b;R_0) \rho_{s} (X_0, t)$.

Applying these projectors to the generalized master equation~(\ref{meqn-wc-formal}) we obtain,
\begin{eqnarray}
   & & \frac{\partial}{\partial t} \mathcal{P}\rho_{d}(X, t) = - \mathcal{P} i L_{d} \mathcal{P} \rho_{d}(X, t) - \mathcal{P} i L_{d} \mathcal{Q} \rho_d(X, t) \nonumber \\
   & & \qquad + \int_0^t dt' \mathcal{P} M (X, t')\mathcal{P} \rho_{d}(\bar{X}_{t'}, t-t')\label{PRME} \\
   & & \quad \qquad + \int_0^t dt' \mathcal{P} M (X, t') \mathcal{Q} \rho_d(\bar{X}_{t'}, t-t') \;,\nonumber\end{eqnarray}
\begin{eqnarray}
   & & \frac{\partial}{\partial t} \mathcal{Q} \rho_d (X, t) = -\mathcal{Q} i L_{d} \mathcal{P} \rho_d (X, t) - \mathcal{Q} i L_{d} \mathcal{Q} \rho_d(X, t) \nonumber \\
   & & \qquad + \int_0^t dt' \mathcal{Q} M (X, t') \mathcal{P}\rho_d(\bar{X}_{t'}, t-t')\label{QRME} \\
   & & \quad \qquad + \int_0^t dt' \mathcal{Q} M (X, t') \mathcal{Q} \rho_d (\bar{X}_{t'}, t-t') \;.\nonumber
\end{eqnarray}
Solving the second equation formally we obtain,
\begin{eqnarray}
   & & \mathcal{Q} \rho_d (X, t) = e^{-i\mathcal{Q} L_d t} \mathcal{Q}\rho_d(X, 0) \label{eq:soln-qrme} \\
   & & \quad - \int_0^t dt' e^{-i \mathcal{Q} L_d t'} i \mathcal{Q}L_d \mathcal{P} \rho_d (X, t - t') + \Phi (X, t) \;,\nonumber
\end{eqnarray}
where the function $\Phi(X, t)$ involves fluctuations of the memory function from its bath average, $\delta M \equiv M - \langle M \rangle_b$, at various time displaced coordinates.  Substituting this solution into Eq.~(\ref{PRME}) gives,
\begin{eqnarray}
   & & \frac{\partial}{\partial t} \rho_s (X_0,t) = \nonumber \\
   & & \quad - \langle i L_d \rangle_b \rho_s (X_0,t) - \int dX_b i L_d e^{-i\mathcal{Q}L_d t} \mathcal{Q}\rho_d (X, 0)\nonumber \\
   & & \quad + \int_0^t dt' \langle i L_d e^{- i \mathcal{Q} L_d t'} i \mathcal{Q} L_d \rangle_b \rho_s (X, t - t') \nonumber \\
   & & \quad + \int_0^t dt' \langle M (X, t') \rangle_b \rho_s (\bar{X}_{t'}, t-t') \label{PRME-folded} \\
   & & \quad + \int_0^{t'} \int dX_b \delta M (X, t') e^{-i \mathcal{Q} L_d (t -
   t')}\mathcal{Q} \rho_d (\bar{X}_{t'}, 0) \nonumber \\
   & & \quad + \int_0^t \int_0^{t - t'}dt' dt'' \langle \delta M (X, t')  e^{- i \mathcal{Q} L_d t''} i \mathcal{Q} L_d \rangle_b  \nonumber \\
   & & \hspace{3.5cm} \times  \rho_s (\bar{X}_{t''}, t - t' - t'') \nonumber \\
   & & \quad + \int_0^t dt' \int dX_b \delta M (X, t') \Phi(\bar{X}_{t'}, t - t') \;.\nonumber
\end{eqnarray}
The last three terms in this equation involve integrals over the bath of expressions containing fluctuations of the memory kernel from its bath average.  These expressions consist of $\delta M$ correlated with dynamical quantities evolved under projected dynamics.  By definition, $\delta M$ is initially zero and, due to the presence of the phase factor, it oscillates strongly for long times. Consequently, the bath integral of the product of the oscillatory function $\delta M$ with a time evolved dynamical quantity is expected to be small.  Taking these considerations into account, we neglect the last three terms in Eq.~(\ref{PRME-folded}). Making this approximation, the subsystem evolution equation takes the form,
\begin{eqnarray}
   & & \frac{\partial}{\partial t} \rho_s (X,t) = \label{PRME-folded-approx1}\\
   & & \quad  - \int dX_b i L_\alpha  e^{-i\mathcal{Q}L_d t} \mathcal{Q}\rho_d (X, 0)  - \langle i L_d \rangle_b \rho_s(X,t) \nonumber \\
   & & \quad - \int_0^t dt' i \langle L_d e^{- i \mathcal{Q} L_d t'} i \mathcal{Q} L_d \rangle_b \rho_s (X, t - t') \nonumber \\
   & &  \quad + \int_0^t dt' \langle M (X, t') \rangle_b \rho_s (\bar{X}_{t'}, t-t') \;.\nonumber
\end{eqnarray}
This equation, written explicitly in terms of its components is given in Eq.~(\ref{penult-PRME-folded}) and forms the basis for the reduction to a master equation.

\end{appendix}


\begin{thebibliography}{53}
\bibitem[{com({\natexlab{b}})}]{com3}
\bibinfo{note}{A quantum-classical description of the system can also be
  derived from considerations based on decoherence by part of the
  system~\cite{shiokawa02}}.
\bibitem[{com({\natexlab{c}})}]{com2}
\bibinfo{note}{For a discussion of the bath projection operator for quantum
  systems, see for example, Romero-Rochin and Oppenheim~\cite{romero-rochin89}.
  For our quantum-classical system we define the projection operator using the
  canonical bath density conditional upon the subsystem coordinate $R_0$.}
\end{thebibliography}
\end{document}